\documentclass[12pt]{article}

\usepackage{graphicx}
\usepackage{amsmath,amssymb}
\usepackage[english]{babel}
\usepackage{cite}
\usepackage[font=footnotesize]{caption}
\usepackage{setspace}

\newcommand{\be}{\begin{equation}}
\newcommand{\ee}{\end{equation}}

\begin{document}
\title{Signatures of HyperCharge Axions at Contemporary and Future Colliders}
\author{\large Ram Brustein, Yael Raveh
\\
\vspace{-.5in}  \vbox{
\begin{center}
$^{\textrm{\normalsize
\ Department of Physics, Ben-Gurion University,
Beer-Sheva 84105, Israel}}$
\\ \small 
ramyb@bgu.ac.il,\ raveh.yl@gmail.com
\end{center}
}}
\date{}

\maketitle

\begin{abstract}
We investigate the possible detection of the Hypercharge Axion (HCA) in colliders. The HCA is a hypothetical pseudoscalar that couples to weak hypercharge topological density and could potentially explain the dominance of matter over antimatter in the observable universe. If the HCA exists, it can be produced in colliders via vector boson fusion or in association with a photon or with a Z boson, and detected by looking for its decay into photons or Z bosons.
We find that for certain values of the HCA mass and coupling scales, both of the order of a TeV, existing data from the Large Hadron Collider (LHC) can already put interesting constraints and in future colliders, such as the High Luminosity LHC, the accessible detection range is increased significantly.
\end{abstract}

\onehalfspacing
\section{Introduction}
Understanding how the baryon asymmetry of the Universe originated is one of the fundamental open questions in High Energy Physics and Cosmology nowadays. Currently, the prevailing theory is that the Universe started out as baryon-symmetric, but now isn't, due to some process that generated matter excess after the Big Bang \cite{Canetti:2012zc}.

One of the suggested baryogenesis mechanisms, incorporates hyperelectric and hypermagnetic fields into electroweak baryogenesis \cite{Anber:2015yca,Kamada:2016eeb}. It was observed that: i) a topological number condensate can be released at the electroweak phase transition (EWPT) in the form of leptons and baryons and ii) strong enough hypermagnetic fields could make the EWPT strongly first-order.

Scalars with axion-like coupling to hypercharge fields were previously considered in \cite{Turner:1987bw,GUENDELMAN1992108}. Amplification of ordinary electromagnetic fields by such scalar fields was discussed in \cite{Turner:1987bw} and their possible use for baryogenesis in \cite{GUENDELMAN1992108}.

More recently, in 2015, the ATLAS and CMS Collaborations at the Large Hadron Collider (LHC) reported preliminary data with a small excess of diphoton events at an invariant mass of about 750 GeV. Though the excess was absent in data collected during 2016 and is now considered a statistical fluctuation \cite{Khachatryan:2016yec,Aaboud:2017yyg}, it nonetheless triggered a lot of attention among the particle physics community; hundreds of theory papers appeared following the 2015 announcement \cite{PhysRevLett.116.150001}. Particularly, several theory papers suggested that if such a discovery was confirmed, it would require unexpected new elementary particles. One of the most widely studied explanations relied on a spin-0 real singlet with effective interactions to the Standard Model (SM) gauge bosons (see, e.g., \cite{Csaki:2015vek,Fichet:2015vvy,Altmannshofer:2015xfo,Harland-Lang:2016qjy}).

The Hypercharge Axion (HCA) is a hypothetical pseudoscalar with electroweak interactions. It was first proposed in 1999 by Brustein and Oaknin \cite{Brustein:1998du} as a candidate for inducing baryogenesis; in cosmology, the HCA can exponentially amplify hypercharge fields in the symmetric phase of the electroweak plasma, while coherently rolling or oscillating \cite{Brustein:1999rk}, leading to the formation of a time-dependent condensate of topological number density. This condensate can be converted at the EWPT, under certain conditions, into baryons in sufficient quantity to explain the observed baryon asymmetry in the Universe \cite{Brustein:1999we}.

Since the hypercharge photon is a linear combination of the ordinary photon and Z boson, HCAs couple to photons and Z bosons. As a result, the HCA can be produced in interactions involving photons and Z bosons and detected by looking for its decay into photons or Z bosons. For this reason, in addition to the fact that the HCA can be relevant to baryogenesis for a range of masses of a  few TeVs \cite{Brustein:1999it}, the HCA Model can be tested  in contemporary and future experiments in particle colliders.

Inspired by the now refuted reports on the 750GeV excess, we initiated a general investigation to determine the possible detection of the HCA in contemporary and future colliders. We followed a standard procedure used most often for phenomenological studies of this kind (see, e.g., \cite{Csaki:2015vek,Fichet:2015vvy,Altmannshofer:2015xfo,Harland-Lang:2016qjy}. In most of these cases, the suggested new particle couples to  additional operators, such as $Y_{\mu\nu}Y^{\mu\nu}$, rather than exclusively to  $Y_{\mu\nu}\widetilde{Y}^{\mu\nu}$ like the HCA. However, the phenomenological procedures are similar to the one performed for the HCA). In our investigation we identified the experimental signatures of the HCA, designed a data analysis strategy that maximizes sensitivity to the HCA Model, and evaluated how much data is required to convincingly establish, or rule out, the model.

This paper is organized as follows: In Sec.~\ref{sec:model} we present a general setup of the HCA Model and discuss theoretical and phenomenological characteristics of the HCA. In Sec.~\ref{sec:prob}, we present criteria for detection and an updated analysis for the possible detection of the HCA at the LHC, as well as at the High Luminosity LHC (HL-LHC) and other future colliders. Some final comments are made in Sec.~\ref{sec:discuss}.

\section{General Setup of the HCA Model} \label{sec:model}
In \cite{Brustein:1999rk}, Brustein and Oaknin discuss hypercharge electrodynamics in the unbroken phase of the electroweak plasma coupled to a cosmological pseudoscalar. We, however, treat below a simpler form of the Lagrangian studied by Brustein and Oaknin; we focus on a singlet elementary HCA whose only coupling to SM fields is to hypercharge fields.

Pseudoscalar fields with axion-like coupling appear in several possible extensions of the SM and typically have only perturbative derivative interactions and therefore vanishing potential. They acquire mass through non-perturbative interactions. Non-perturbative effects generate a potential of the form $V(\phi)=V_0^4V(\phi/f)$, where $V$ is a bounded periodic function characterized by the mass generation scale, $f$, also known as the ``Peccei-Quinn'' scale. The scale $f$ could be as high as the Planck scale or much lower, even down to the TeV range.

By coupling the HCA, $X$, to hypercharge electromagnetic fields, $Y_{\mu \nu}$, and considering for simplicity $V(X/f) = \Lambda^4 (1-\cos{X/f})$, we find that the SM Lagrangian is supplemented by:
\be
\label{eq:lag} L=\dfrac{1}{2}(\partial_{\mu}X)^2-\Lambda^4 \left( 1-\cos{\dfrac{X}{f}}\right) -\dfrac{1}{4M}\epsilon^{\mu \nu \rho \sigma}XY_{\mu \nu}Y_{\rho \sigma}.
\ee
Note:

$\bullet$ The HCA can be displaced from its minimum in the early universe with interesting consequences, but for now we assume that it has reached its global minimum at $X = 0$, where its mass is given by $m_X\equiv \Lambda^2/f \ll M$. The RHS of the last expression ensures small radiative corrections (see below).
\vspace{0.2cm}

$\bullet$ The mass acquired by the HCA could be as low as a fraction of an eV, or as high as $10^{12}$ GeV. A particularly interesting mass range is the TeV range, expected to appear if mass generation is associated with supersymmetry breaking and if the HCA plays a role in baryogenesis \cite{Brustein:1999rk}.
\vspace{0.2cm}

$\bullet$ The scales M in the hypercharge sector and $\Lambda$, $f$ in the mass generation sector, are not related. Our model is therefore a two parameters model, and the goal of our analysis is to determine for which domains in ($m_X,M$) space, HCA can be produced and detected in colliders.

\begin{figure}[t]
\includegraphics[scale=1,trim={4.1cm 21.05cm 4cm 4cm},clip]{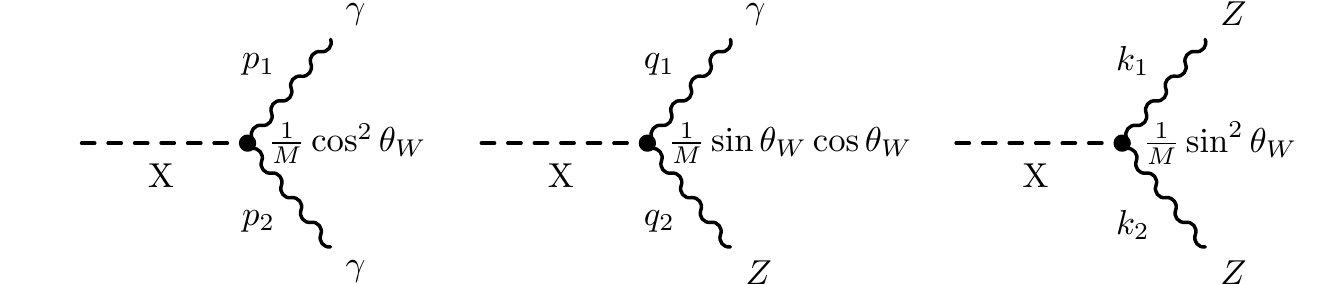}
\caption{The HCA couplings to photos and Z bosons.} \label{fig:coupling}
\end{figure}
\paragraph*{}
The part of the Lagrangian that is of interest for detection of a HCA is the last term in Eq.~(\ref{eq:lag}). We have decomposed the interaction term into the ordinary photon and Z boson and obtained the couplings shown in Fig.~\ref{fig:coupling}. Next, we calculated partial decay widths, and these are the expressions obtained
\be \label{eq:br} \begin{split}
&\Gamma_{X\rightarrow \gamma \gamma}=\dfrac{1}{64\pi M^2}\cos^4{\theta_W}[m_X^3] \\ &\Gamma_{X\rightarrow Z\gamma}=\dfrac{2}{64\pi M^2}\cos^2{\theta_W}\sin^2{\theta_W}\left[ \dfrac{(m_X^2-m_Z^2)^3}{m_X^3}\right]  \\ &\Gamma_{X\rightarrow ZZ}=\dfrac{1}{64\pi M^2}\sin^4{\theta_W}[(m_X^2-4m^2_Z)^{3/2}]. \\ &\Gamma=\Gamma_{\gamma \gamma}+\Gamma_{Z \gamma}+\Gamma_{ZZ}.
\end{split} \ee
The branching ratios of a HCA whose mass ranges few TeVs depend weakly on its mass, $m_X$, and do not depend on the coupling $1/M$. Moreover, we found that $\Gamma_{\gamma\gamma}/\Gamma$ dominates throughout the range of parameters researched.

\subsection{Phenomenological Methods} \label{sec:ph}
Throughout this study, events were generated in \texttt{M{\small AD}G{\small RAPH}5\textunderscore {\footnotesize A}MC@NLO}, a Monte Carlo event generator used most often for simulating particle colliders \cite{Alwall:2014hca}. Cross sections for production processes of the HCA were evaluated using a simulation code that fits the Brustein and Oaknin proposed theory; the effective operator in Eq.~(\ref{eq:lag}) had been implemented\footnote{Specifically for a HCA of mass $m_X=750$ GeV, and $M=100$ TeV. The coupling and parameters adjustments required for our analysis are easy to make using \texttt{M{\scriptsize AD}G{\scriptsize RAPH}5\textunderscore {\tiny A}MC@NLO}.} in model UFO files created with \texttt{FeynRules} \cite{Alloul:2013bka}.

We have conducted several validation tests on \texttt{M{\small AD}G{\small RAPH}5\textunderscore {\footnotesize A}MC@NLO} with well known cross sections in perturbation theory, as well as with processes which involve the HCA. All validation tests were successful; numerical simulations reproduced complete perturbative results to O(.1\%).

\subsection{Production Mechanisms}
\begin{figure*}[t]
\includegraphics[scale=1,trim={3.7cm 20.95cm 4cm 4cm},clip]{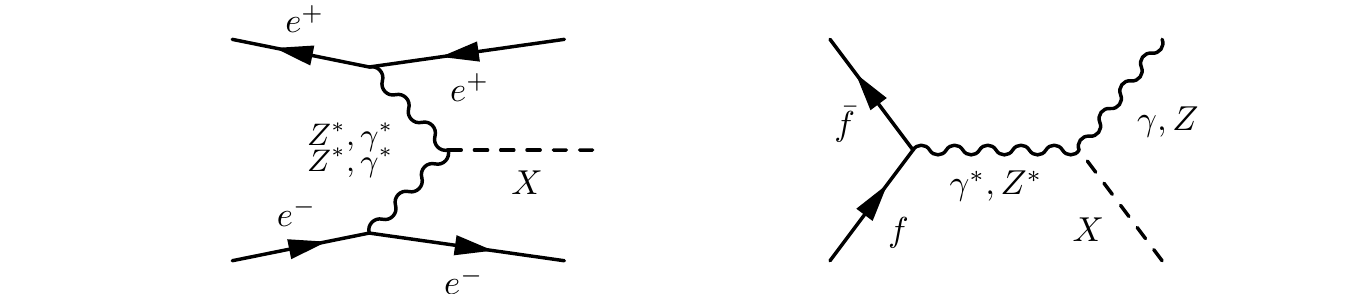}
\caption{Production of the HCA via (left) VBF and (right) AP. $Z^*,\gamma^*$ are virtual bosons. The fermions can be either charged leptons or quarks. At hadron colliders, the VBF-induced HCA production is treated as a double deep-inelastic scattering (see below).} \label{fig:production}
\end{figure*}

The HCA can be produced at high energy colliders via vector boson fusion (VBF) or via associated production (AP) with another photon or Z boson. Both VBF and AP are depicted in Fig.~\ref{fig:production}.

The AP channel had been studied by Brustein and Oaknin \cite{Brustein:1999it}. Using the AP diagrams shown in Fig.~\ref{fig:production}, we calculated the total unpolarized cross sections for the two processes $f\overline{f}\rightarrow Z^*,\gamma^* \rightarrow ZX$ and $f\overline{f}\rightarrow Z^*,\gamma^* \rightarrow \gamma X$ and obtained
\be \label{eq:apa} \begin{split} \sigma(f\overline{f}\rightarrow XZ)=&\dfrac{\alpha}{12N_c}\dfrac{1}{M^2}\sqrt{\hat{s}}E_z^3\rho^3\dfrac{\sin^2{\theta_w}}{\cos^2{\theta_w}}\times \\ &\left\lbrace \left[ \dfrac{c_V}{\hat{s}-m_z^2}+2\dfrac{\cos^2{\theta_w}}{\hat{s}}(g_A/e)\right] ^2+\dfrac{c_A^2}{(\hat{s}-m_z^2)^2}\right\rbrace \end{split} \ee
\be \label{eq:apz} \begin{split} \sigma(f\overline{f}\rightarrow X\gamma)=&\dfrac{\alpha}{12N_c}\dfrac{1}{M^2}\sqrt{\hat{s}}E_{\gamma}^3\times \\ &\left\lbrace \left[ \dfrac{c_V}{\hat{s}-m_z^2}+2\dfrac{\cos^2{\theta_w}}{\hat{s}}(g_A/e)\right] ^2+\dfrac{c_A^2}{(\hat{s}-m_z^2)^2}\right\rbrace \end{split} \ee
where $E_Z=\frac{m_Z^2-m_X^2+\hat{s}}{2\sqrt{\hat{s}}}$, $\rho =\sqrt{1-\frac{m_Z^2}{E_Z^2}}$, $E_{\gamma}=\frac{\hat{s}-m_X^2}{2\sqrt{\hat{s}}}$, $N_c$ is the number of colors of the fermion which takes into account averaging over initial colors and $\sqrt{\hat{s}}$ denotes the parton center of mass (CM) energy. In deriving Eqs.~(\ref{eq:apa}) and (\ref{eq:apz}) we assumed that the fermions are effectively massless, $m_f\ll \sqrt{\hat{s}}$. The parameters appearing in Eqs.~(\ref{eq:apa}) and (\ref{eq:apz}) are given in Table \ref{tab:1}: $c_V$ and $c_A$ are the vector and axial coupling of the fermion to the Z boson.
\begin{table}[b]
\begin{center}
    \begin{tabular}{| l | c | c | c |}
    \hline & & & \\[-0.8em]
    \textbf{$f$} & \textbf{$g_A$} & \textbf{$c_V$} & \textbf{$c_A$} \\ [0.2em] \hline & & & \\[-0.8em]
    $e^-, \ \mu^-, \ \tau^-$ & $e=\sqrt{4\pi \alpha}$ & $-\frac{1}{2}+2\sin^2{\theta_w}$ & $-\frac{1}{2}$ \\ [0.8em]
    $u, \ c, \ t$ & $-\frac{2}{3}\sqrt{4\pi \alpha}$ & $\frac{1}{2}-\frac{4}{3}\sin^2{\theta_w}$ & $\frac{1}{2}$ \\ [0.8em]
    $d, \ s, \ b$ & $\frac{1}{3}\sqrt{4\pi \alpha}$ & $-\frac{1}{2}+\frac{2}{3}\sin^2{\theta_w}$ & $-\frac{1}{2}$ \\ [0.4em]
    \hline
    \end{tabular}
    \caption{Parameter values for fermions. $\alpha$ denotes the fine structure constant.}\label{tab:1}
    \end{center}
    \end{table}

Notice that for small $\hat{s}$, there are kinematical thresholds for both processes, $\hat{s} > m_X^2$ and $\hat{s} > (m_X +m_Z)^2$, to allow AP with a photon and with a Z boson, respectively. Additionally, since $E_{\gamma},E_Z\sim \sqrt{\hat{s}}$ for large $\hat{s}$, both cross sections approach asymptotically a constant [apart from a logarithmic dependence of $\alpha(\sqrt{\hat{s}})$] independent of the mass $m_X$, such that for large $\hat{s}$, $\dfrac{\sigma(f\overline{f}\rightarrow ZX)}{\sigma(f\overline{f}\rightarrow \gamma X)}\simeq \tan^2{\theta_W}\simeq 0.3$. The rise towards the asymptotic value is governed by the ratios $m_X^2/\hat{s}$ and $m_Z^2/\hat{s}$.

In order to evaluate cross sections at hadron colliders such as the LHC, we used the parton model. For both AP processes, the only possible contributions are from the 6 same-flavour quark-antiquark partonic collisions, whose cross sections are given in Eqs.~(\ref{eq:apa}) and (\ref{eq:apz}).

\begin{figure}[t]
\includegraphics[scale=0.413,trim={1.8cm 7.15cm 2.8cm 7.6cm},clip]{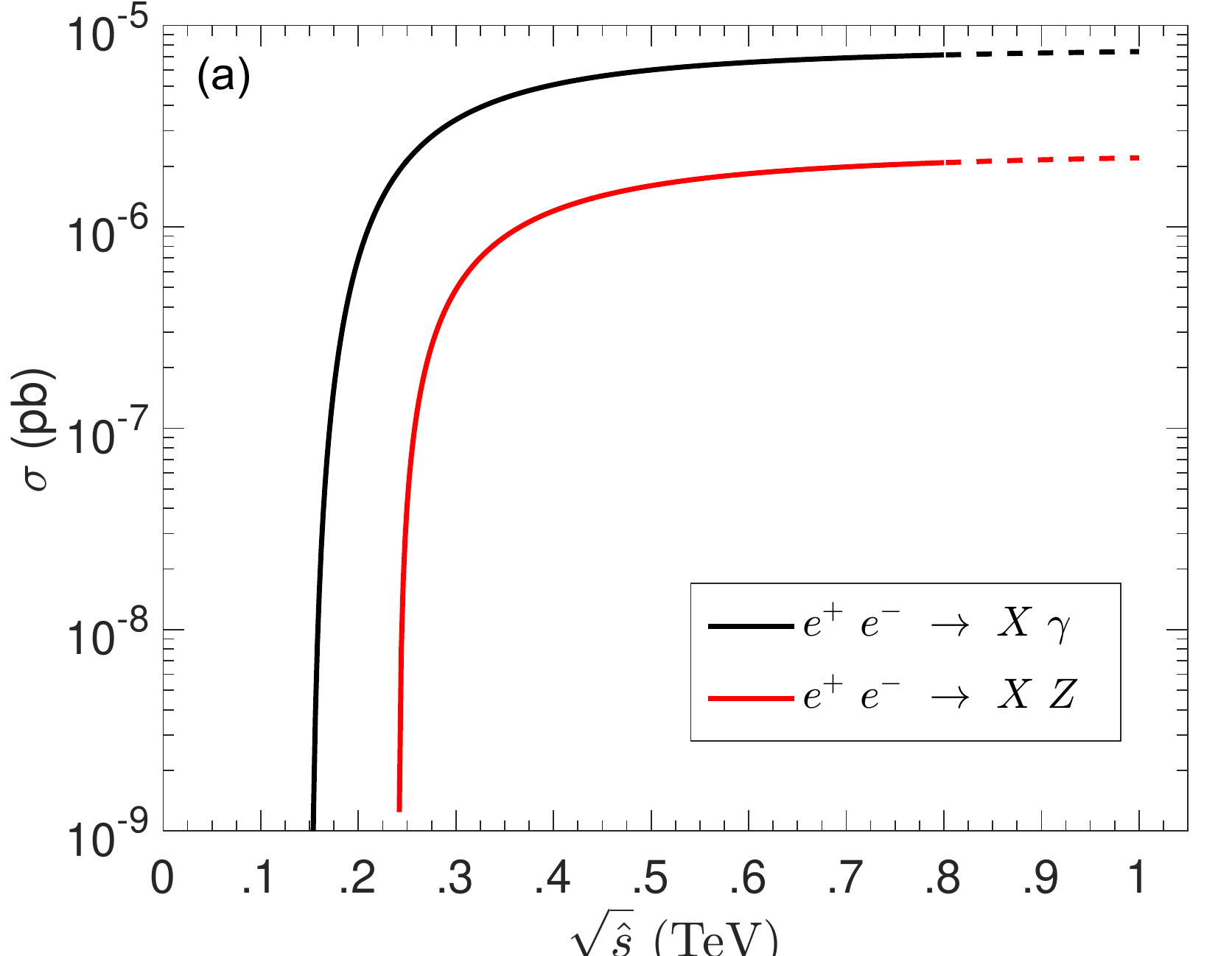}\includegraphics[scale=0.4,trim={1.55cm 6.9cm 2.6cm 7.35cm},clip]{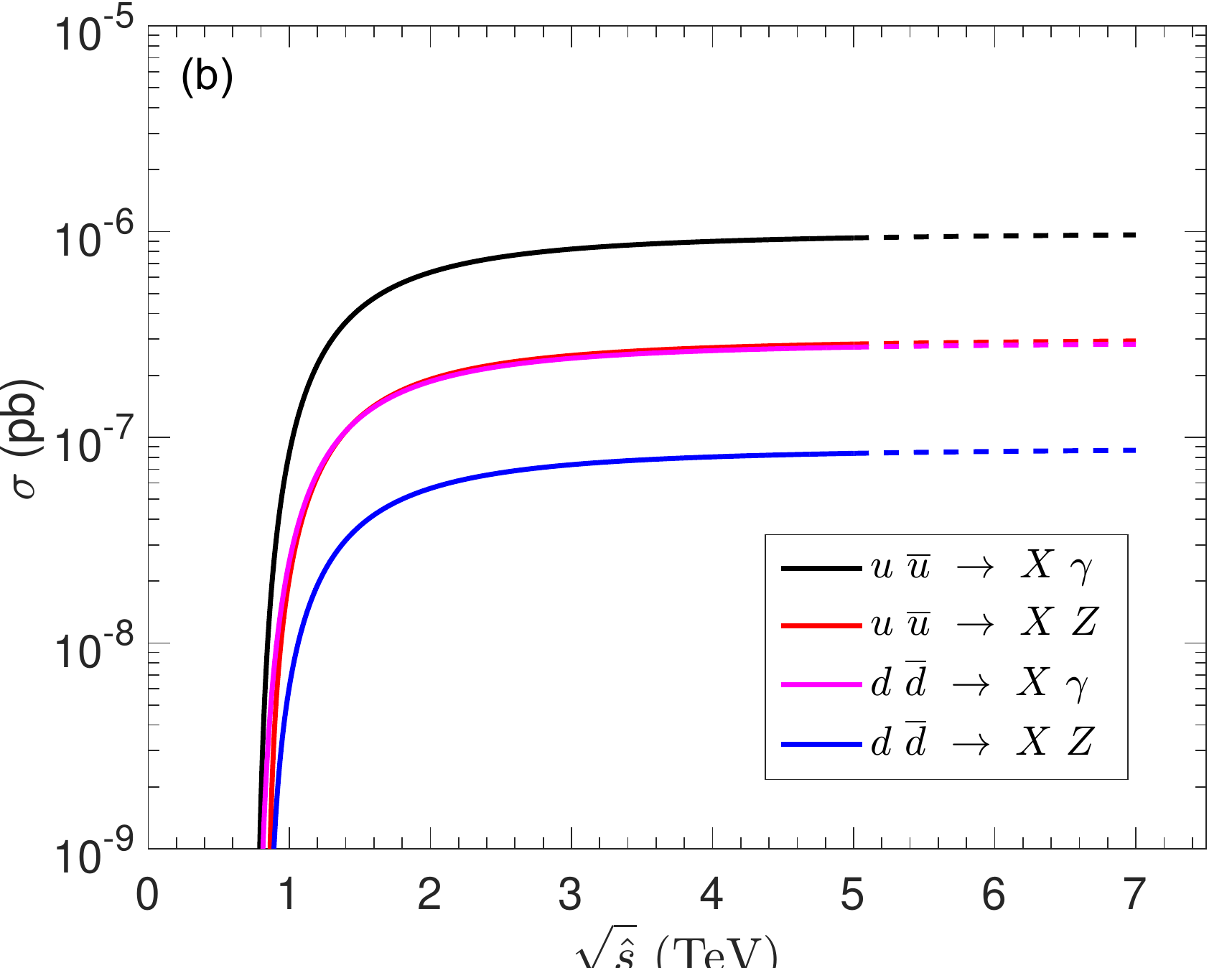}
\caption{Cross sections for AP processes as a function of the partons CM energy at (a) $e^+e^-$ colliders, for $m_X=150$ GeV,$M=100$ TeV and (b) hadron colliders, for $m_X=750$ GeV, $M=100$ TeV.}\label{fig:APgraphs}
\end{figure}

The AP cross-sections calculated using perturbation theory, Eqs.~(\ref{eq:apa}) and (\ref{eq:apz}), were checked against \texttt{M{\small AD}G{\small RAPH}5\textunderscore {\footnotesize A}MC@NLO} results. The software uses constantly-updated PDFs of protons in order to simulate pp collisions.
In Fig. \ref{fig:APgraphs} we present the AP cross-sections as a function of the parton CM energy at (a) $e^+e^-$ colliders, for $m_X=150$ GeV and $M=100$ TeV, and at (b) hadron colliders for $m_X=750$ GeV and $M=100$ TeV.

The VBF mechanism and possible detection of HCAs at the ATLAS detector were studied by Elfgren \cite{Elfgren:2000ch}. An analytic expression of cross sections for VBF processes, unfortunately is not very illuminating as it depends on the vector-boson distribution functions in the colliding fermions. Nonetheless, in order to get a rough estimate of the processes cross sections we use the Weizsacker-Williams approximation \cite{vonWeizsacker:1934nji}, also known as the equivalent photon approximation (EPA). The EPA seems to give a reasonable order of magnitude estimate of the cross section. The EPA estimate becomes better for CM energies that are much higher than $m_X^2$.

The Weizsacker-Williams photon spectrum, $f_{f/\gamma}(x)$, is the photon distribution with momentum fraction $x$ in a charged-particle beam of energy $E$. For an electron, or a light quark of energy $E$, the probability of finding a collinear photon of energy $xE$ is given by
\be \label{eq:wiz} f_{f/\gamma}(x)=\dfrac{g_A^2}{8\pi^2}\dfrac{x+(1-x)^2}{x}\ln{\left( \dfrac{t_{max}}{t_{min}}\right) }. \ee
Here $t_{max}$ and $t_{min}$ are the characteristic maximum and minimum photon momentum transfers and $g_A$ is given in Table \ref{tab:1}. For the process under consideration, the production of heavy HCA of mass $m_X$, we take these to be $t_{min} = 1$ GeV$^2$ and $t_{max} = \hat{s}$, with $\hat{s}$ being the partonic CM energy. There is some flexibility in the choice of $t_{max}$. However, the results are not very sensitive to this parameter within the limits of the Weizsacker-Williams approximation \cite{Bhattacharya:1995id}. The particular choice of the minimum momentum transfer, $t_{min}$, guarantees that the photons are obtained from the deep inelastic scattering of protons and the quark-parton model is valid.

One could generalized EPA to what is known as the effective vector-boson approximation (EVBA) \cite{DAWSON198542,Kane:1984bb} for processes with weak bosons in place of photons \cite{Bhattacharya:1995id}. That, however, can not be done analytically and is beyond the scope of this study. In addition, the production of a HCA whose mass is within the range researched of few TeVs, is dominated by the exchange of two photons. We therefore content ourselves with HCA production via two photons only and obtain an expression for the parton-level cross section
\be
\begin{split} &\sigma^{inel}(\hat{s})\approx\int{dxf_{f/\gamma}(x)\int{dyf_{f/\gamma}(y)\sigma(\gamma \gamma \rightarrow X,xy\hat{s})}} \\ &=\int_0^1{dxf_{f/\gamma}(x)\int_0^1{dyf_{f/\gamma}(y)\dfrac{\langle \vert \mathcal{M}\vert^2\rangle}{2(2xE)(2yE')}2\pi\delta((k+k')^2+m_X^2)}}.
\end{split}
\ee
Here $\mathcal{M}$ denotes the invariant amplitude for production of the HCA in a collision of two photons with momenta $k,k'$, in a charged-particle beam of energies $E,E'$ respectively.

\begin{figure*}[t]
\includegraphics[scale=0.405,trim={1.65cm 6.85cm 2.45cm 7.35cm},clip]{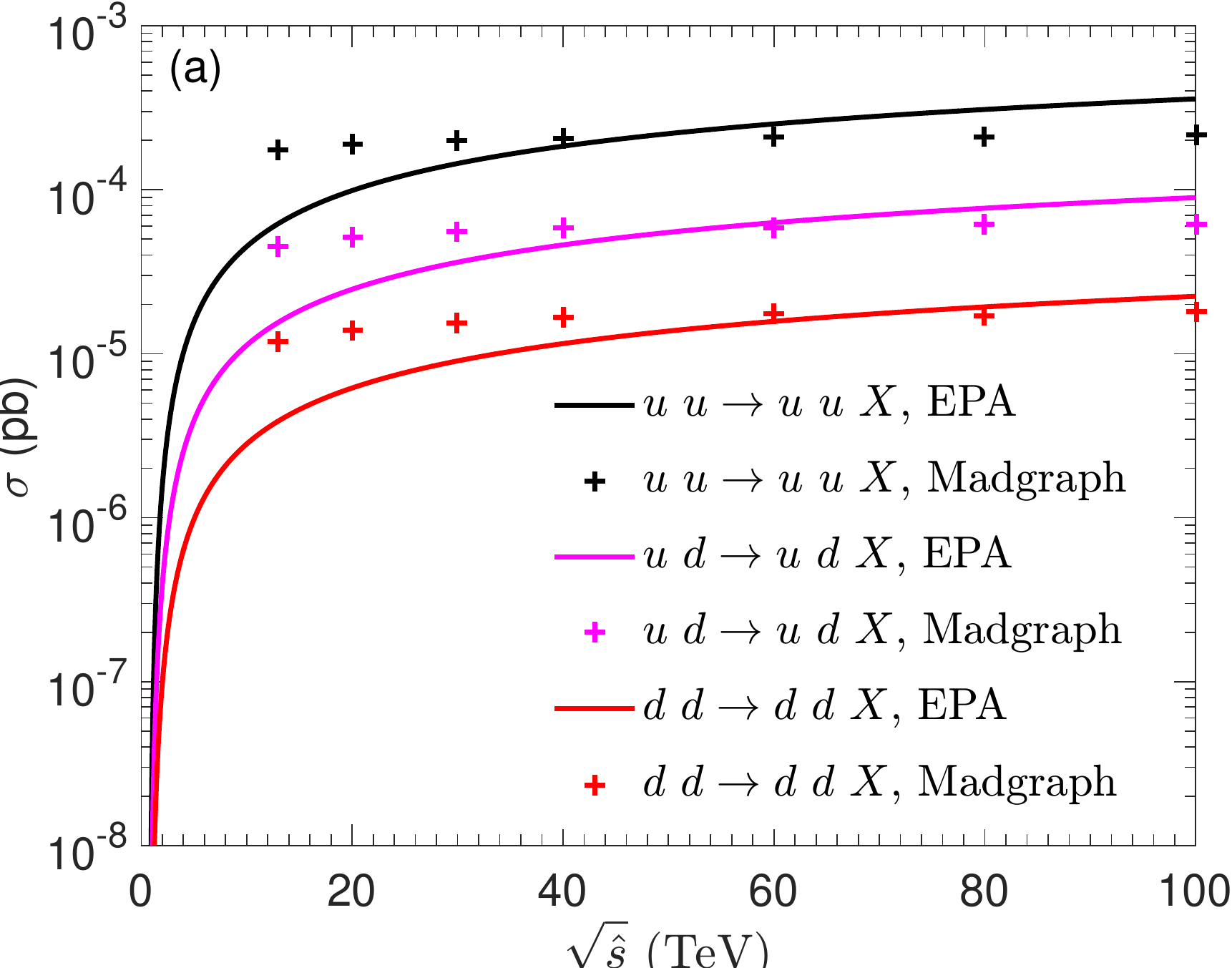}
\includegraphics[scale=0.395,trim={1.6cm 6.85cm 2.1cm 7.25cm},clip]{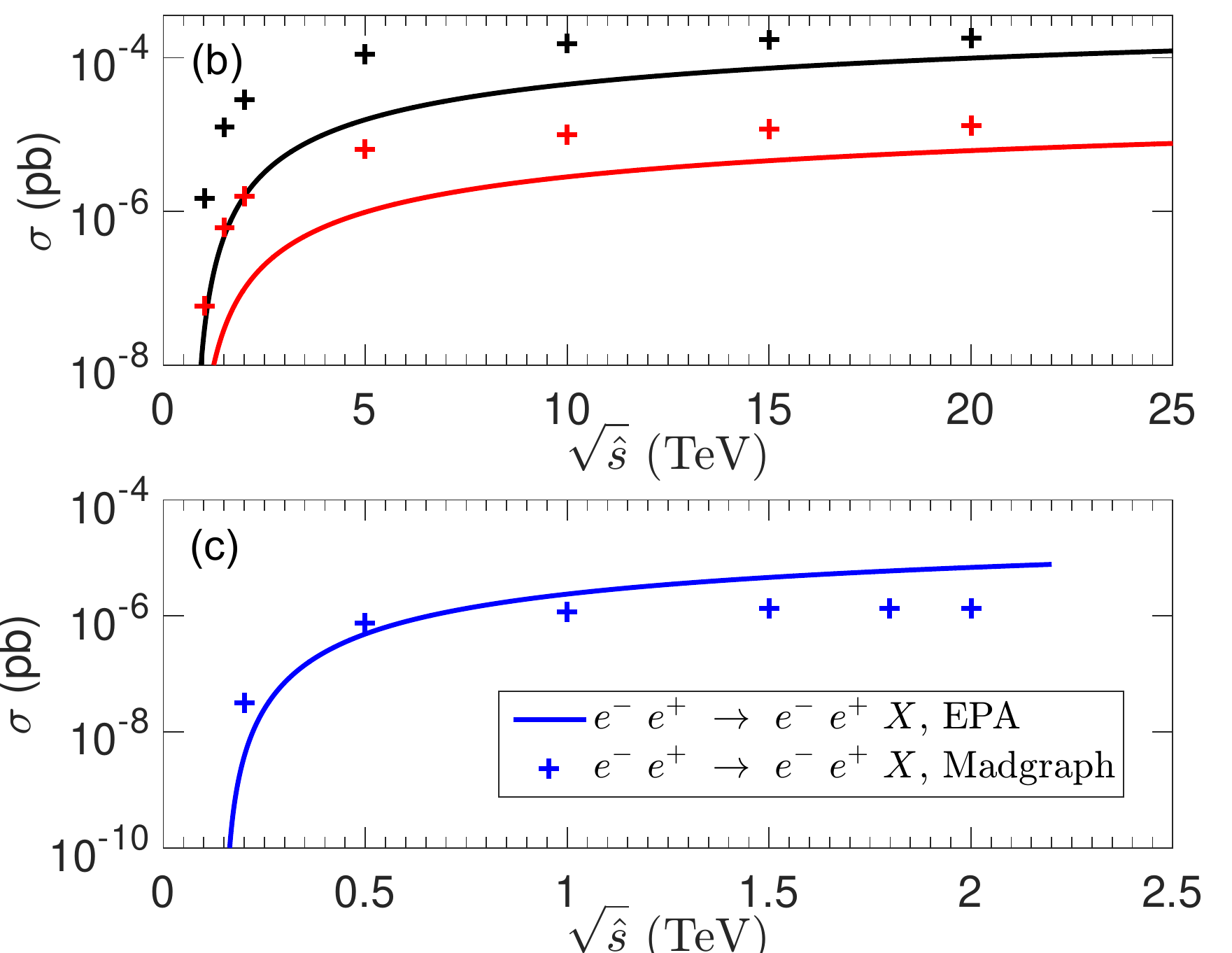}
\caption{Parton-level cross sections for VBF processes at (a-b) hadronic colliders with $m_X=750$ GeV, $M=10$ TeV and (c) $e^+e^-$ colliders with $m_X=150$ GeV, $M=100$ TeV vs. partons CM energy. For clarity, only the leading VBF processes at pp colliders are presented. The solid lines are valid to all light quarks of the type specified in the legend. EPA estimates correctly the order of magnitude of the perturbative results  only for CM energies much larger than $m_X^2$. For $m_Z^2\ll \hat{s}$ one has to consider contributions from production processes via $Z\gamma$ and $ZZ$.}
\label{fig:VBFgraphs}
\end{figure*}

Integrating over $y$ using the $\delta$-function, averaging over initial colors and expressing $\langle \vert\mathcal{\overline{M}}\vert^2\rangle$ in terms of $\Gamma(X\rightarrow \gamma\gamma)$, we find
\be \begin{split}
\sigma^{inel}(\hat{s})&=\dfrac{16\pi^2\Gamma(X\rightarrow \gamma\gamma)}{m_X\hat{s}N_c}\int_{m_X^2/\hat{s}}^1 {\frac{dx}{x}f_{f/\gamma}(x)f_{f/\gamma}(m_X^2/x\hat{s})} \\ &\equiv \dfrac{16\pi^2\Gamma(X\rightarrow \gamma\gamma)}{m_X\hat{s}N_c}\dfrac{dL_{ff/\gamma\gamma}}{d(m_X^2/\hat{s})}.
\end{split}
\ee
An analytic expression can be obtained for the differential luminosity. Thus, by substituting  $\Gamma(X\rightarrow \gamma\gamma)$ [Eq.~(\ref{eq:br})], we obtain
\be
\label{eq:vbf}
\begin{split}
&\sigma^{inel}(\hat{s})\approx  \dfrac{\pi\alpha^2}{256 \hat{s}N_c}\left( \frac{g_A}{e}\right) ^4\frac{m_X^2}{M^2}\cos^4{\theta_W}\times \\ &
\left[ \left( 4-\frac{6\hat{s}}{m_X^2}+\frac{2m_X^2}{\hat{s}}\right) -
\left( 4+\frac{4\hat{s}}{m_X^2}+\frac{m_X^2}{\hat{s}}\right) \ln{\frac{m_X^2}{\hat{s}}}\right]
\left( \ln{\frac{t_{max}}{t_{min}}}\right) ^2.
\end{split}
\ee
The approximation is only valid when the transverse momentum of the fermions is virtually zero and when they are ultra-relativistic. This should be a fairly good approximation, as the fermions are either high energy electrons in $e^+e^-$ colliders, or the quarks inside a proton which have very little transverse momentum and a relativistic velocity.

In Fig.~\ref{fig:VBFgraphs} we present cross-sections for production of the HCA via VBF as a function of $\hat{s}$ at hadronic and $e^+e^-$ colliders. We see that EPA can reproduce the order of magnitude of the complete perturbative results. The approximation is valid for CM energies much larger than $\hat{s}\sim m_X^2$. For very large CM energy ($m_Z^2\ll \hat{s}$) one has to consider contributions from production via $Z\gamma$ and $ZZ$.
\begin{figure*}[t]
\includegraphics[scale=0.4,trim={1.6cm 6.75cm 2.4cm 6.7cm},clip]{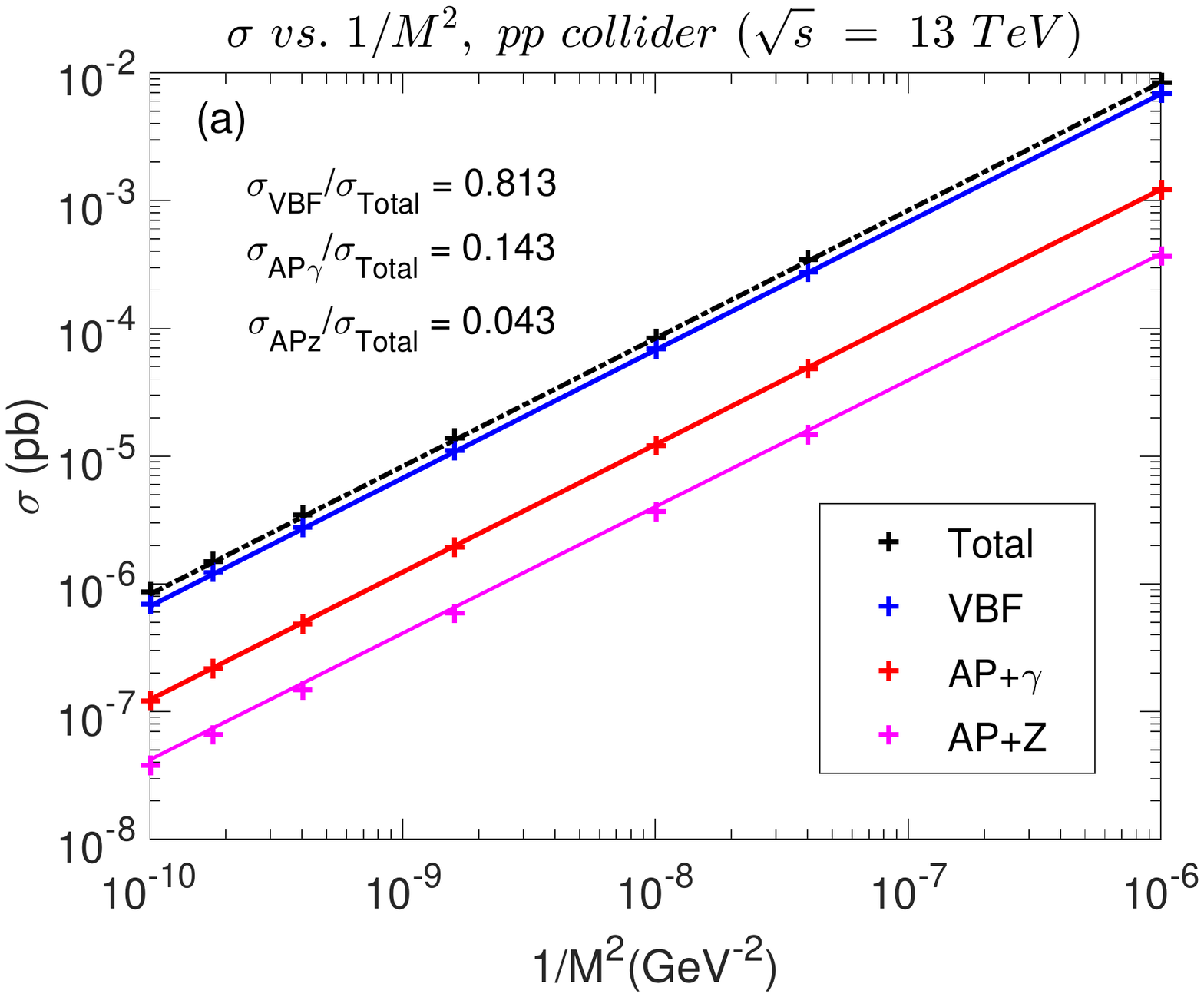}\includegraphics[scale=0.4,trim={1.6cm 6.75cm 2.4cm 6.7cm},clip]{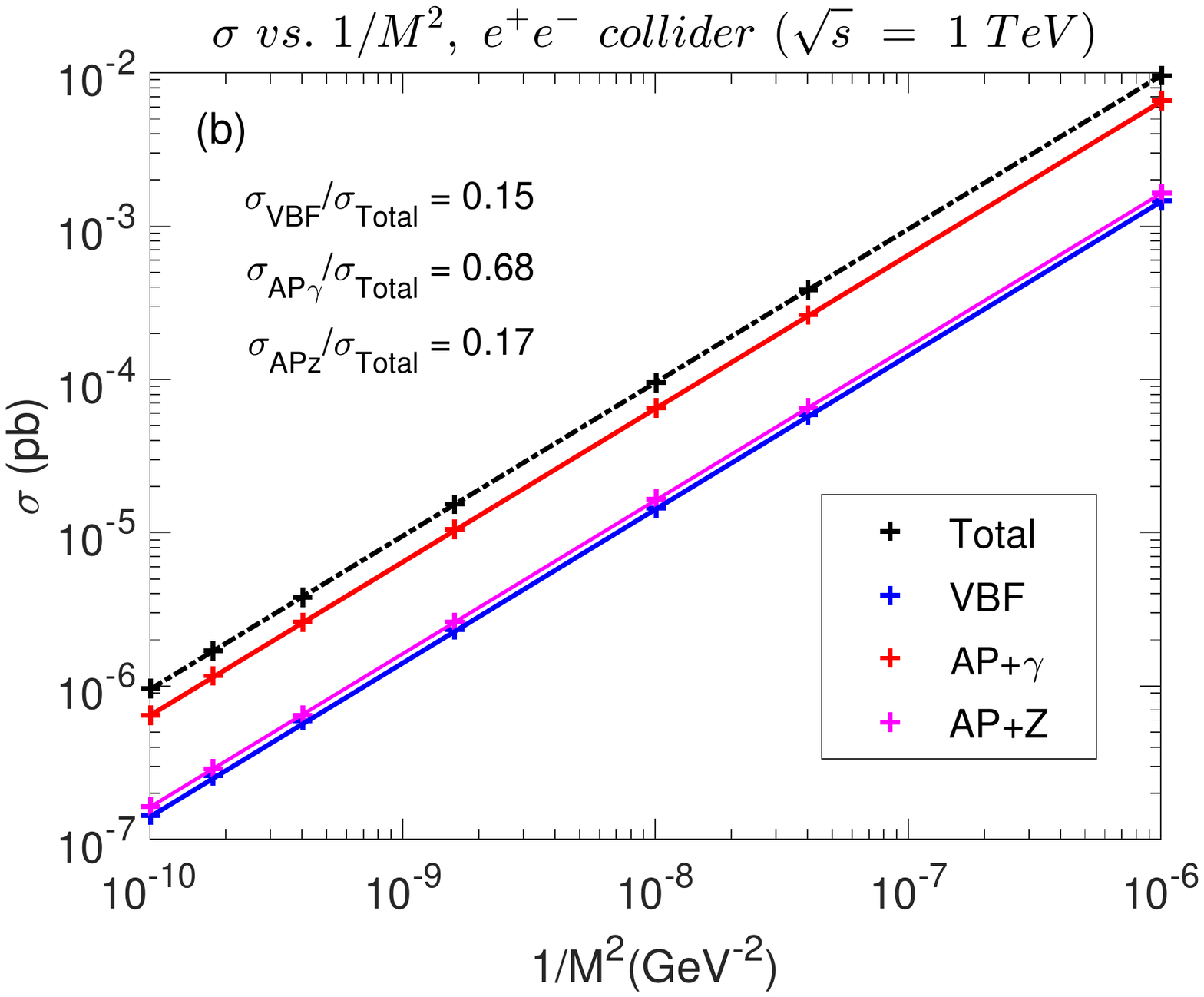} \caption{HCA production ($m_X=750$ GeV) at (a) the LHC, and (b) a next generation $e^+e^-$ collider. All lines are fitted curves. VBF is the most probable production process at hadron colliders, whereas at $e^+e^-$ colliders, the production of the HCA in association with a photon is favourable.}\label{fig:favorable}
\end{figure*}
\begin{figure*}[t]
\includegraphics[scale=0.41,trim={1.85cm 6.8cm 2.5cm 6.7cm},clip]{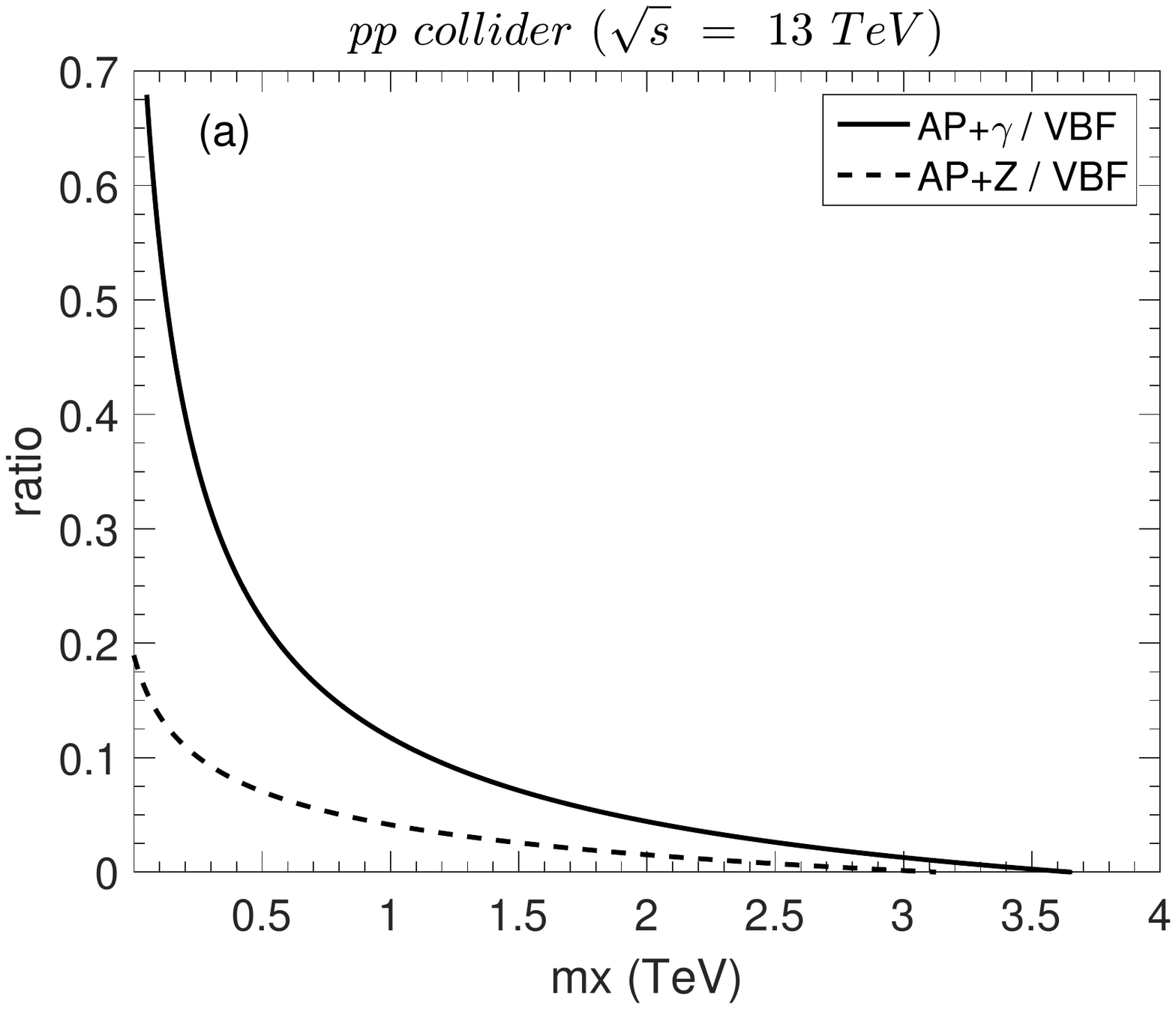}\includegraphics[scale=0.41,trim={1.85cm 6.8cm 2.5cm 6.7cm},clip]{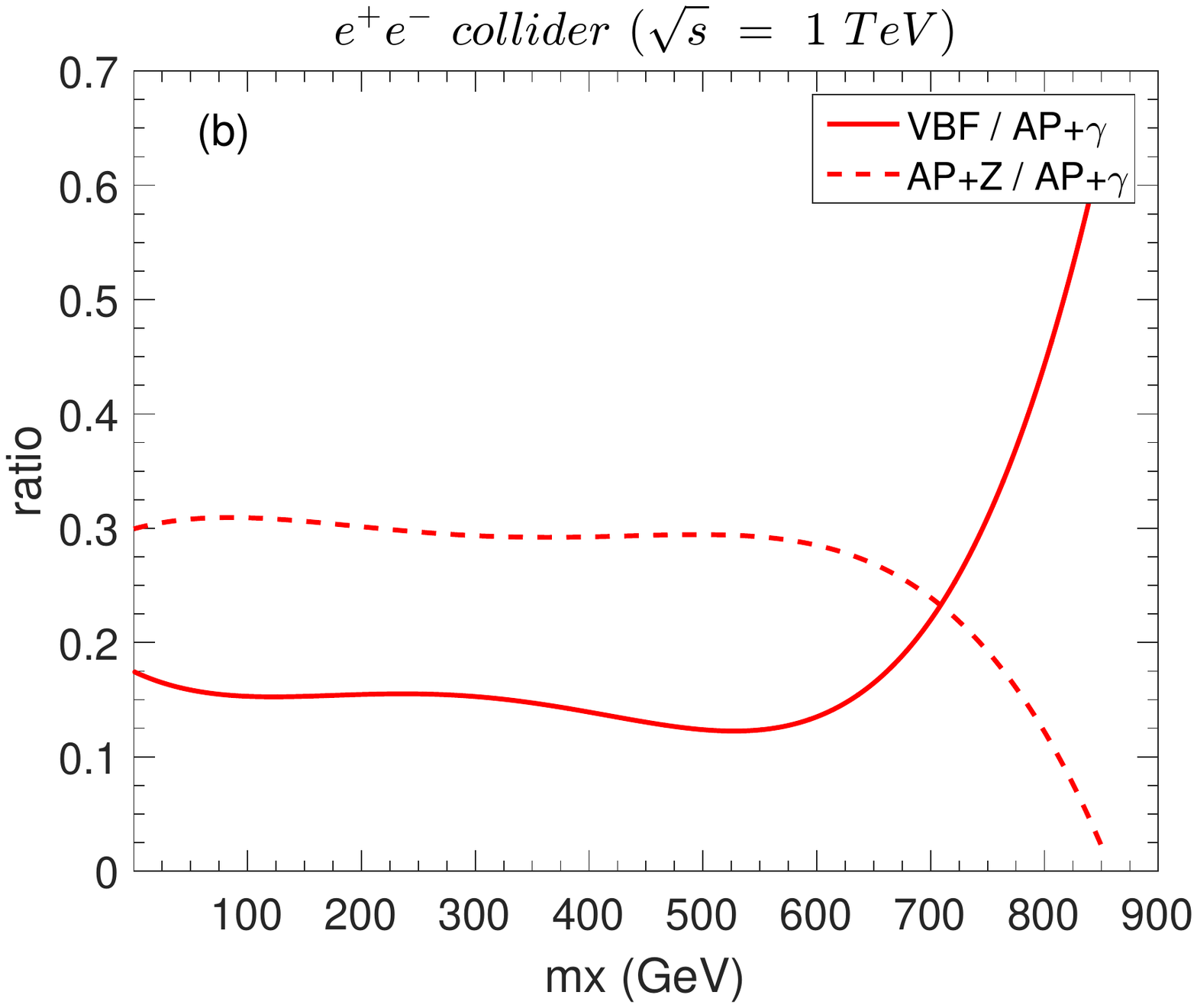} \caption{(a) Ratios of cross sections for AP of HCA with a photon and with a Z boson, to cross section for production via VBF at the LHC. (b) Ratios of cross sections for AP of HCA with a Z boson and production via VBF, to cross section for AP with a photon at a next generation $e^+e^-$ collider. In both cases $M=10$ TeV.}\label{fig:favorable1}
\end{figure*}
\paragraph*{}
According to Figs.~\ref{fig:favorable} and \ref{fig:favorable1}, the most probable production mechanism of the HCA at hadron colliders is the VBF, whereas at $e^+e^-$ machines, the production of the HCA in association with another photon is favorable.

\subsection{Decay Channels}
Assuming that $m_X \geqslant 2m_Z$, all vertices in the HCA Model (Fig.~\ref{fig:coupling}) are kinematically allowed. The HCA can therefore decay via all three channels. Eqs.~(\ref{eq:br}) suggest that  widths in the GeV range can be expected for $m_X\lesssim M\sim$ TeV. If there is a larger hierarchy between the scales, the width can be much smaller.

Theoretically, large widths can be accommodated \cite{Ben-Dayan:2016gxw}. That being said, narrower width gives a ``healthier'' model; the interaction term is not renormalizable and therefore our model should be considered as an effective field theory, with a cutoff. Loop corrections could thus present a problem, or even invalidate our analysis, if they are as large as the tree level ones. This concern is certainly relevant in the case $m_X\sim M$, as the relative magnitude of a loop with external gauge bosons and an internal HCA, compared to the same process at tree level is approximately of order $\frac{1}{16\pi^2}\frac{m_X^2}{M^2}$, assuming that the HCA is the heaviest particle running in the loop \cite{Brustein:1999it}. However, if we manage to stay within a realm where $m_X < M$, expected radiative corrections are small. Moreover, we recall that the HCA potential is generated by non-perturbative effects. As such, naive dimensional analysis does not necessarily hold. We will not attempt here a more detailed treatment of radiative corrections.

According to Figs.~\ref{fig:favorable} and \ref{fig:favorable1}, the dominant production mechanism depends on the collider type. Therefore, for the sake of generality, we characterize the HCA using its decay products only. Having said that, decay products of a VBF-induced HCA production are expected to be emitted back-to-back, generally with high $p_T$. In addition, the AP key signature is, as well, a very distinctive event with a three photons final state. The SM background for the three photons final state at $e^+e^-$ colliders is mainly a pure QED process, whose differential cross-section is strongly peaked along the forward and backward directions \cite{Devoto:2005pd,BERENDS198122}, whereas the three photons coming from near resonance AP and decay of the HCA were found to be isotropically distributed. These properties can be used to distinguish the HCA signal from background events, thus serve as very powerful checks for verifying discovery.

To characterize the HCA final decay products at particle detectors, we examine the Z boson decay products\cite{Patrignani:2016xqp}. The Z decays to two leptons of each flavor 3.3\% of the time, to two quarks 70\% of the time and into neutrinos the rest of the time. The $\tau$ is more difficult to detect at the LHC and is not considered here, leaving the electron and the muon for a total of 6.7\% of the detectable lepton decays. We thus obtain the HCA final decay products with their respective ratios that are shown in Table~\ref{tab:br}.

The final decay products of the HCA appear with fixed ratios and can therefore serve as a powerful diagnostic for verifying discovery of the HCA; if several different final states are detected in the predicted ratios, the new particle could be identified as the HCA.
\begin{table}[t]
\begin{center}
    \begin{tabular}{| c | c | c | c |}
    \hline & & & \\[-0.8em]
    \textbf{Decay} & \textbf{Branching} & \textbf{Detectable} & \textbf{Branching} \\
    \textbf{modes} & \textbf{Ratio} & \textbf{final state} & \textbf{Ratio}\\
    [0.2em] \hline & & & \\[-0.8em]
    $\gamma\ \gamma$ & 59\% & $\gamma\ \gamma$ & 100\% \\ [0.2em] \hline & & & \\[-0.8em]
    $\gamma\ Z$ & 35\% & $\gamma~l~\overline{l}$ & 6.7\% \\
     & & $\gamma~q~q$ & 70\% \\ [0.2em] \hline & & & \\[-0.8em]
    $Z\ Z$ & 6\% & $l~\overline{l}~l~\overline{l}$ & 0.45\% \\
     & & $l~\overline{l}~q~q$ & 9.4\% \\
     & & $q~q~q~q$ & 49\% \\ [0.2em]
    \hline
    \end{tabular}
    \caption{Decay modes of the HCA, branching ratios and detectable final states.}\label{tab:br}
    \end{center}
    \end{table}

Since $X\rightarrow Z+Z$ is rare, and $X\rightarrow \gamma +jet+jet$ is very polluted by the QCD background at hadron colliders, the most promising final states for detection are:
\be
\begin{split}
&X \rightarrow \gamma +\gamma, \\
&X \rightarrow \gamma +Z \rightarrow \gamma +l+\overline{l}.
\end{split}
\ee

\section{Detection at Present and Future Colliders} \label{sec:prob}
Finally, we present the estimated number of produced HCAs in colliders, $N_{ev}$. Eqs. (\ref{eq:apa}), (\ref{eq:apz}) and (\ref{eq:vbf}), along with the expression $N_{ev}=\mathcal{L}_{int}\cdot \sigma$, where $\mathcal{L}$ is the integrated luminosity of the experiment, relate $N_{ev}$ and the two parameters of the model; the HCA mass and the coupling.
\begin{table}[t]
\begin{center}
    \begin{tabular}{| l | c | c | c | c |}
    \hline & & & & \\[-0.8em]
    \textbf{Name} & \textbf{Type} & \textbf{$\mathcal{L}_{int}$ (fb$^{-1}$)} & \textbf{$\sqrt{s}$ (TeV)} & \textbf{Years of} \\ & & & & \textbf{operation} \\ [0.2em] \hline & & & & \\[-0.8em]
    LHC I & & $23.3$ at $8$ TeV & 7 - 8 & 2010-2013 \\
     & & $6.1$ at $7$ TeV & & \\
    LHC II & pp & $80$ & $13$ & 2015-now \\
     & & $120 - 150$ * & $13$ & 2015-2018 \\
    LHC III & & $300$ * & $14$ * & 2020-2023 \\ [0.2em]  \hline & & & & \\[-0.8em]
    HL-LHC & pp & $250$/y * & $14$ * & 2026-2038? \\ [0.2em] \hline & & & & \\[-0.8em]
    ILC & $e^+e^-$ & $1.5\times 10^{-5}$/s * & $0.5-1$ * & TBD \\ [0.2em] \hline & & & & \\[-0.8em]
    FCC-hh & pp & $0.2-1$ M/y * & $100$ * & ??? \\ [0.2em] \hline
    \end{tabular}
    \end{center}
    \caption{Table of parameter values for high-energy colliders \cite{Patrignani:2016xqp}. The parameters expected at the LHC experiments for the 2018 run and the design values for the high-luminosity upgrade (HL-LHC) are also shown. Tentative design parameters of selected future high-energy colliders are denoted with a *.}\label{tab:coll}
    \end{table}

If we fix a number of events as the minimal number $N_{min}$ required for detection of the HCA, we obtain an ``accessible detection range'' in ($m_X,M$) space, limited by the curve
\be
\label{eq:sig} \frac{M}{TeV}=\sqrt{\frac{\mathcal{L}_{int}}{N_{min}}}\sqrt{\sigma(M=1TeV,s,m_x)}. \ee
\begin{figure}[p]
\vspace{.1in}
\begin{center}
\includegraphics[scale=0.565,trim={1.95cm 5.88cm 1.7cm 7.45cm},clip]{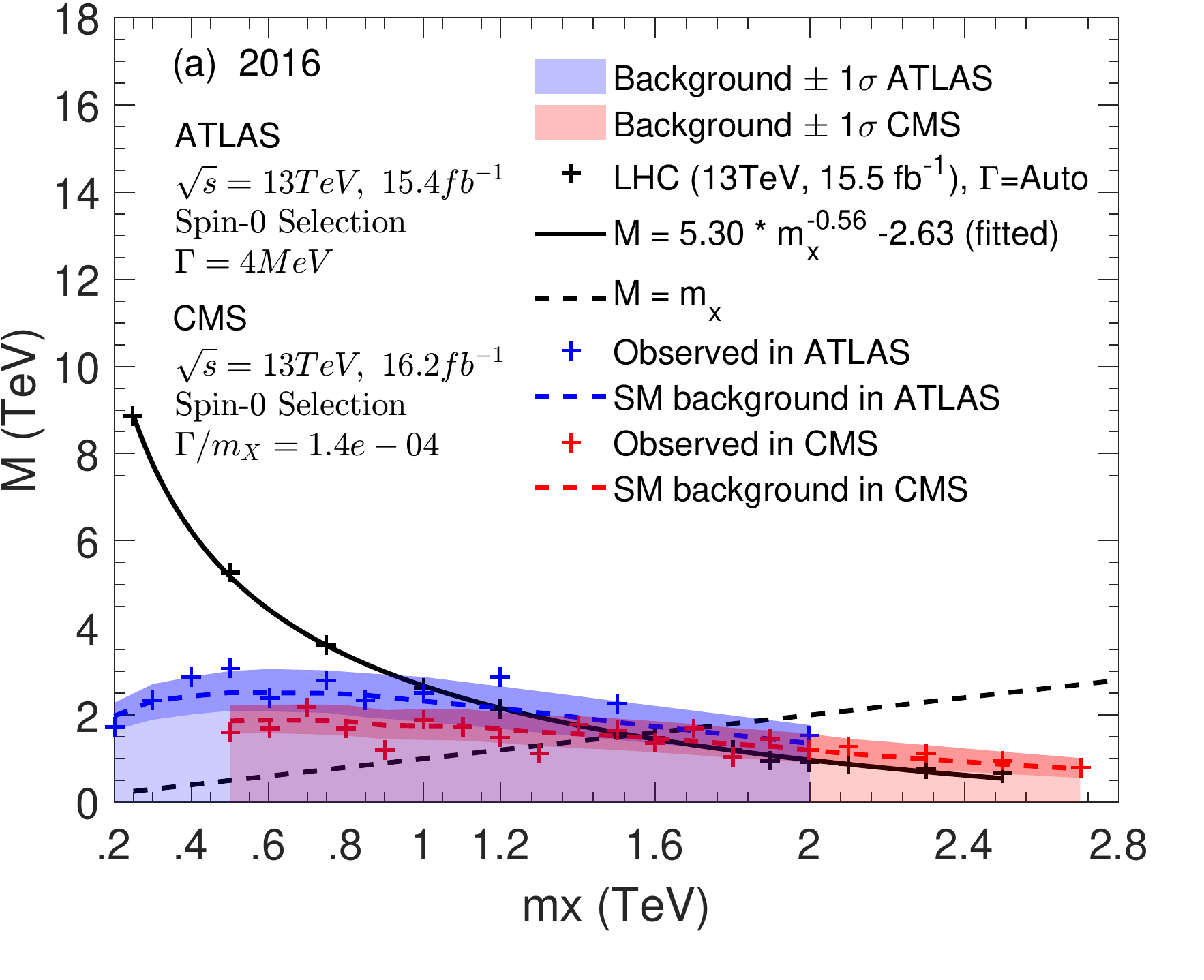}
\includegraphics[scale=0.565,trim={1.65cm 6.25cm 1.4cm 7.25cm},clip]{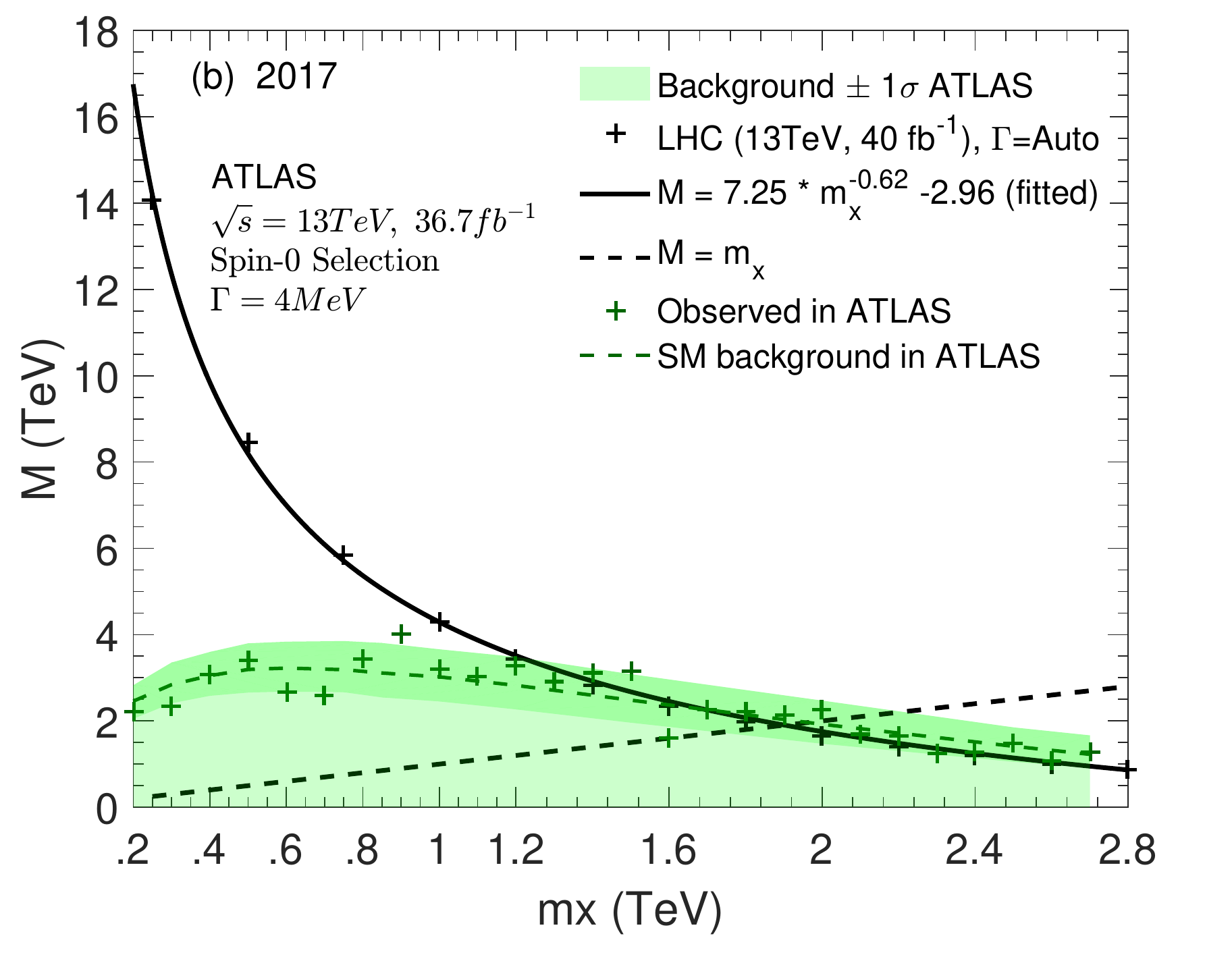}
\caption{Expected number of HCAs produced in Run II of the LHC; $\sqrt{s}=13$ TeV (a) $\mathcal{L}_{int}=15.5$ fb$^{-1}$, collected with the LHC detectors during 2015 and 2016, and (b) $\mathcal{L}_{int}=40$ fb$^{-1}$, collected with the ATLAS detector by the end of 2017. The shaded regions are excluded for different decay widths. Note that different cuts were used in the event selection procedures of ATLAS and CMS analyses (\cite[sections 4-7]{Aaboud:2017yyg,ATLAS:2016eeo} and \cite{Khachatryan:2016yec}, respectively). The solid lines correspond to 10 expected events.}
\label{fig:exclusion}
\end{center}
\end{figure}
In Fig.~\ref{fig:exclusion}, this curve is shown for $N_{min}=10$ for Run II of the LHC, along with an exclusion region obtained from previous experiments (see below). The reach of future colliders, amongst which are future runs of the LHC, HL-LHC and a FCChh, is shown in Fig.~\ref{fig:detect}. For the graphs in Figs.~\ref{fig:exclusion} and \ref{fig:detect}, we have chosen some generic parameters characterizing the various machines, summarized in Table \ref{tab:coll}.

In the region below the curves more than 10 events are expected with luminosities given in legends, while in the region above the curves less than 10 events are expected. Since graphs scale with parameters as in Eq.~(\ref{eq:sig}), it is simple to adapt them to different parameters characterizing the various machines.

Our choice of $N_{min}=10$ events is motivated by the expectation that the SM background for the processes is small. The actual reach of future colliders for detection of the HCA will be determined by the background.

\paragraph{}
The HCA had not been detected so far. Nonetheless, we can use existing experimental data from recent particle experiments to rule out significant regions in the model's parameters space. In order to do so, we use the analyses in \cite{ATLAS:2016eeo,Khachatryan:2016yec} and the more recent one in \cite{Aaboud:2017yyg}. These analyses summarize the results of recent searches for high-mass diphoton resonances with different widths, specifically spin-0 and spin-2 resonances with an invariant mass between hundereds of GeVs to few TeVs in pp collisions at a CM energy of $13$ TeV. The data samples correspond to an integrated luminosity of $15.4$ fb$^{-1}$ collected with the ATLAS detector and $16.2$ fb$^{-1}$ collected with CMS during 2015 and 2016, and $36.7$ fb$^{-1}$ collected with the ATLAS detector by the end of 2017 (\cite{ATLAS:2016eeo,Khachatryan:2016yec} and \cite{Aaboud:2017yyg} respectively). No significant excess had been observed relative to the SM expectation.

Since HCAs within the ``accessible detection range'' in ($m_X,M$) space, or ($m_X\sim 1$ TeV, $M\sim 10$ TeV) according to Fig.~\ref{fig:detect}, have narrower decay width of few MeVs, we further examine the graphs which correspond to limits on the signal cross section times branching ratio to two photons for a spin-0 particle as a function of the assumed signal mass and for narrow-width resonance, as in Fig.~7(a) of \cite{ATLAS:2016eeo}, Fig.~4(upper) of \cite{Khachatryan:2016yec} and Fig.~4(a) of \cite{Aaboud:2017yyg}. First, we calculated the total cross section per mass value by dividing the aforementioned plots with the branching ratio to two photons in the HCA Model. Then, using \texttt{M{\small AD}G{\small RAPH}5\textunderscore {\footnotesize A}MC@NLO}, we calculated $M$ values.

In their study, Brustein and Oaknin analyzed data from Tevatron and LEP II which set upper bounds on the cross sections $\sigma(e^+e^-\rightarrow XZ)$, of the order of $0.1$ pb, for $m_X < 90$ GeV and assuming that $X \rightarrow \gamma \gamma$ \cite{Brustein:1999it}. The updated analysis we performed, summarized in Fig.~\ref{fig:exclusion}(b), clearly enlarges the exclusion region obtained by Brustein and Oaknin, thus provides stricter constraints on regions in ($m_X,M$) space in which the HCA can be produced and detected at the LHC. That being said, a specific analysis with particular details of the HCA Model, should be carried out in order to determine the exact exclusion region.
\begin{figure}[p] 
\includegraphics[scale=0.65,trim={0.3cm 6.7cm 0.5cm 6.8cm},clip]{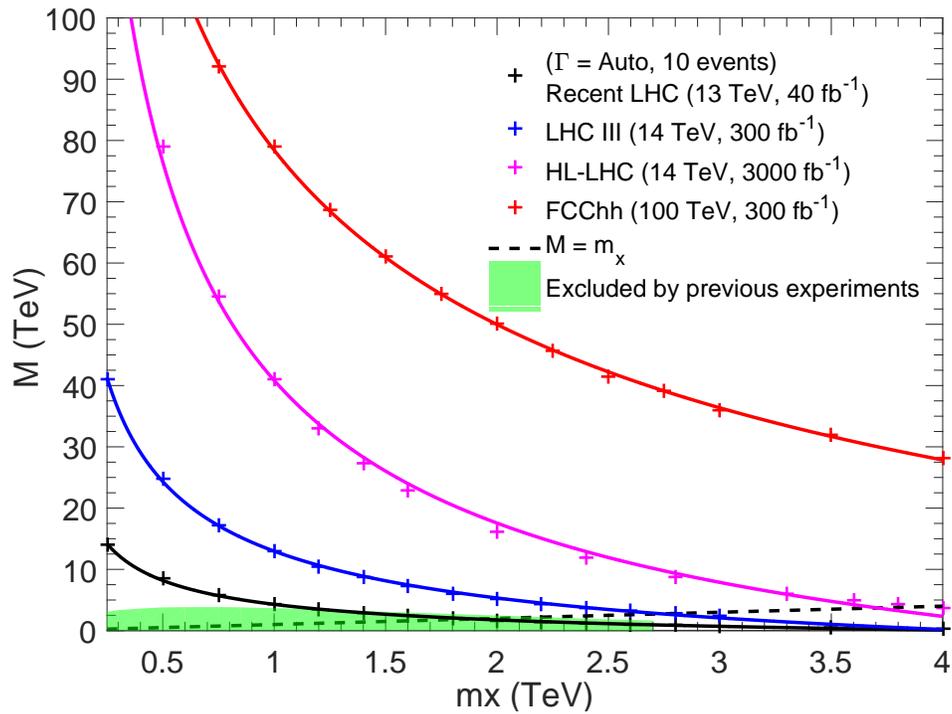}
\caption{Expected number of HCAs produced in future runs of the LHC, HL-LHC and about one year of a FCChh. The solid lines are fitted curves which correspond to 10 expected events; below (above) the curve more (less) than 10 events are expected, with integrated luminosity given in the legend. The shaded region is experimentally excluded.}
\label{fig:detect}
\end{figure}

The main results of our analysis concern future colliders and are summarized in Fig.~\ref{fig:detect}. Firstly, we conclude that for the range of masses $m_X\lesssim 1$ TeV, the HCA could be detected even for the $80$ fb$^{-1}$ accumulated luminosity available today at a CM energy of $13$ TeV. Such large accessible detection region calls, in our opinion, for further experimental attention.

In addition, our results indicate that the accessible detection region increases dramatically as the luminosity and the CM energy increase. Quantitatively one can say that since $N_{sig}/\sqrt{N_{SM}}\propto\sqrt{\mathcal{L}_{int}}$, the HL-LHC is expected to increase the accessible detection range by a factor of $\sqrt{10}$ beyond the accessible region at the LHC by the end of its operational period. Moreover, though there aren't any near-future practical intentions of building higher than the LHC energy colliders, we have obtained quite an interesting result that while a luminosity increase provides better reach in coupling $M$, higher collision energy was found able to significantly increase the reach in mass, $m_X$, as well.

Before concluding our analysis, it is important to note that though a clear signal for the $X \rightarrow \gamma\gamma$ could support a discovery and verify the branching ratios, it is not enough for discovery. If we want to be certain that it is really the HCA that is detected, the branching ratios of  $X \rightarrow \gamma\gamma$ and  $X \rightarrow \gamma Z \rightarrow \gamma l\overline{l}$, the second most promising signature, have to be compared. The ratio $\Gamma_{\gamma\gamma}/\Gamma_{\gamma l \overline{l}}$ should be 25.1 for the HCA. If all three decay channels are discovered, then it is a strong indication that we have discovered a (pseudo)scalar with the right coupling to hypercharge. At that stage, the mass-scale $M$ can be used to determine theoretically whether the pseudoscalar could amplify the hypermagnetic fields or not \cite{Brustein:1999we}.

\section{Conclusions} \label{sec:discuss}

Our study focuses on the HCA Model; a model that could potentially serve as a viable baryogenesis mechanism. The model has great testability potential by contemporary and future experiments. Our general approach has been illustrated in detail, for what is probably the minimal extension of the SM in this context; the addition of a single (pseudo)scalar field that specifically couples to weak hypercharge.

The hypercharge photon is a linear combination of the ordinary photon and $Z$-boson, hence the hypothetical particle could be produced, in hadronic or leptonic colliders, through VBF, or in association with another photon or Z-boson. Figs. \ref{fig:favorable} and \ref{fig:favorable1} show that the VBF process is the most promising production process of the HCA at hadron colliders, whereas at $e^+e^-$ colliders, the production of the HCA in association with another photon is the most probable process.

The produced HCA decays into two neutral gauge bosons, $\gamma\gamma$, $\gamma Z$ or $ZZ$. The experimental signatures of these processes are, then, two or three neutral gauge bosons produced in well defined ratios. We concluded that there are practically two processes that can be used for detection of the HCA; $X \rightarrow \gamma\gamma$ and $X \rightarrow \gamma Z \rightarrow \gamma l \overline{l}$. The former signal is expected to have the higher signal-to-background ratio. Moreover, we argued that the almost isotropic angular distribution of momenta of the outgoing bosons could help in separating it from the QED background at $e^+e^-$ colliders.

Our analysis updates a similar analysis performed by Brustein and Oaknin two decades ago, to contemporary and future colliders. It suggests that the accessible HCA parameter region for the detection of the HCA at future colliders increases dramatically as the luminosity and CM energy increase; higher effective luminosity increases the reach in coupling, while higher energy increases the reach in mass. In addition, our analysis significantly broadens the exclusion region obtained by Brustein and Oaknin, thus provides stricter constraints on domains in the model's parameter space in which the HCA can be produced and detected at colliders.

An additional important implication of our results is that there are parameter regions where the potential of detecting the HCA has not yet been exhausted. We find that even for the $80$ fb$^{-1}$ accumulated luminosity available today at a CM energy of $13$ TeV, the region of parameter space of relatively small mass is still not excluded. This calls, in our opinion, for further experimental attention. One could additionally use recent reports by ATLAS and CMS Collaborations, \cite{Aaboud:2017uhw,Khachatryan:2016odk}, in order to study the second most ``detectable'' signature, $X \rightarrow \gamma Z \rightarrow \gamma l \overline{l}$, and use it to help in verifying branching ratios in particular and a discovery in general.

Due to lack of experimental data, the analysis presented here unfortunately does not cover certain necessary aspects of a meaningful search, which may be easily integrated and complement our research, such as systematic uncertainties elimination, background estimations at future colliders, and more. Hopefully, our study will inspire particle experimenters, or a future study in general, to further examine these and other aspects of the HCA Model in more detail.

\begin{center} \textbf{Acknowledgments} \end{center}
We would like to thank Liron Barak, Yevgeny Kats and David Oaknin  for many helpful discussions. The research was supported by the Israel Science Foundation grant no. 1294/16.

\end{document}